\RequirePackage{fix-cm}
\documentclass[smallextended]{svjour3}       
\smartqed  
\usepackage{graphicx}
\usepackage[numbers]{natbib}
\usepackage{physics}
\usepackage{amsfonts}
\usepackage{subfig}
\usepackage{float}
\usepackage{amsmath}
\usepackage[labelfont=bf]{caption}
\usepackage{url}

\usepackage{breakurl}
\usepackage[breaklinks]{hyperref} 
\usepackage{hyperref}
\usepackage{academicons}
\usepackage{xcolor}

\def\makeheadbox{\relax}

\newcommand{\orcid}[1]{\href{https://orcid.org/#1}{\textcolor[HTML]{A6CE39}{\aiOrcid}}}

\begin{document}

\title{Quantum Measurement Classification with Qudits
}

\titlerunning{Quantum Measurement Classification with Qudits}        

\author{Diego H. Useche \and Andres Giraldo-Carvajal\and Hernan M. Zuluaga-Bucheli \and Jose A. Jaramillo-Villegas 
 \and Fabio A. González 
}


\institute{Diego H. Useche \at
              MindLab Research Group, Universidad Nacional de Colombia, Bogotá, Colombia \\
              \email{diusecher@unal.edu.co}           
          \and
          Andres Giraldo-Carvajal \at
              Facultad de Ingenierías, Universidad Tecnológica de Pereira, Pereira, Colombia \\
              \email{andregiraldo@utp.edu.co}  
          \and
          Hernan M. Zuluaga-Bucheli \at
              Facultad de Ingenierías, Universidad Tecnológica de Pereira, Pereira, Colombia  \\
              \email{herzulu@utp.edu.co}  
          \and
          Jose A. Jaramillo-Villegas \at
              Facultad de Ingenierías, Universidad Tecnológica de Pereira, Pereira, Colombia  \\
              Laboratory for Research in Complex Systems, Menlo Park, California, USA.\\
              \email{jjv@utp.edu.co}  
            \and
          Fabio A. González \at
              MindLab Research Group, Universidad Nacional de Colombia, Bogotá, Colombia \\
              \email{fagonzalezo@unal.edu.co} 
}

\date{}

\maketitle

\begin{abstract}
This paper presents a hybrid classical-quantum program for density estimation and supervised classification. The program is implemented as a quantum circuit in a high-dimensional quantum computer simulator. We show that the proposed quantum protocols allow to estimate probability density functions and to make predictions in a supervised learning manner. This model can be generalized to find expected values of density matrices in high-dimensional quantum computers. Experiments on various data sets are presented. Results show that the proposed method is a viable strategy to implement supervised classification and density estimation in a high-dimensional quantum computer. 

\keywords{Quantum Computing \and Qudit \and Quantum Machine Learning \and Quantum Measurement Classification \and High-dimensional quantum computing}
\end{abstract}



\section{Introduction}

Quantum computing has gained a lot of attention in recent years due to its potential to solve complex problems which would take exponential time in classical computers. Most of the research efforts have been focused on constructing quantum computers based on qubits \cite{Arute2019QuantumProcessor}. However, there has been a growing interest in building quantum computers based on qudits, i.e. machines that simulate and operate d-dimensional quantum states, with $d > 2$. Various physical implementations of high-dimensional quantum states have been proposed, such as photonic states integrated in chips \cite{Schaeff2015ExperimentalOptics,Carolan2015UniversalOptics}, photonic modes encoded in the orbital angular momentum (OAM) \cite{Sit2017High-dimensionalPhotons}, ion traps \cite{Klimov2003QutritIons}, ququarts implemented on a quadrupolar nuclear magnetic resonance (NMR) \cite{Gedik2015ComputationalQudit}, and molecular quantum magnets \cite{Moreno-Pineda2018MolecularAlgorithms}. Two of the main advantages of high-dimensional quantum computers compared to their qubit-based counterparts are their larger information storage \cite{Cozzolino2019High-dimensionalChallenges}, and their higher resilience to noise \cite{Sheridan2010SecuritySystems}. 

One closely related field of quantum computing is quantum machine learning (QML). This field aims to develop novel quantum-inspired machine learning (ML) methods that may run on classical or quantum computers and to implement the existing ML algorithms on quantum computers. For instance, some classical machine learning algorithms like support vector machines and restricted Boltzmann machines can be implemented on qubit-based quantum computers \cite{Rebentrost2014QuantumClassification, Wiebe2014QuantumLearning}, and many of the ML methods have been reformulated in the language of quantum physics like quantum decision trees \cite{Lu2014QuantumClassifier}, quantum neural networks \cite{Ezhov2000QuantumNetworks, Cong2019QuantumNetworks}, and quantum generative adversarial networks \cite{Dallaire-Demers2018QuantumNetworks}. In contrast with QML methods built on qubits, less research has been done on QML based on qudits, i.e. algorithms that run in high-dimensional quantum computers. Some of these methods include protocols with qudits for reinforcement learning \cite{Cardenas-Lopez2018MultiqubitTechnologies}, and for training quantum neural networks \cite{Diep2020SomeNetworks, NIPS2003_50525975, Beer2019EfficientNetworks}.

In addition to the aforementioned methods, Gonzalez et. al. \cite{Gonzalez2021}, proposed two quantum-inspired machine learning methods, the Density Matrix Kernel Density Estimation (DMKDE), which is a non-parametric density estimation method, and the Density Matrix Kernel Density Classification (DMKDC), a supervised machine learning algorithm based on density matrices and kernel density estimation. In this article, we propose two quantum protocols to implement the prediction phase of the DMKDE and the DMKDC in a high-dimensional quantum computer, the simulations were performed using the high-dimensional quantum simulator QuantumSkynet \cite{giraldocarvajal2021quantumskynet}. 


This article is organized as follows: In Section 2, we present the background with the descriptions of the DMKDE, the DMKDC, and the quantum simulator QuantumSkynet with some of its quantum gates, in Section 3, we describe the proposed high-dimensional quantum circuits, in Section 4, we show some results of the method on some toy data sets, and finally, we present the conclusions of the work in Section 5.

\section{Background}

In this section, we present a review of the Density Matrix Kernel Density Estimation (DMKDE), and the Density Matrix Kernel Density Classification (DMKDC) methods proposed by Gonzalez et. al. \cite{Gonzalez2021}, which are the basis of this article. In addition, we describe the high-dimensional quantum computer simulator, QuantumSkynet, with some qudit-based quantum gates which were applied in this work. 
  
\subsection{Density Matrix Kernel Density Estimation (DMKDE)}

The Density Matrix Kernel Density Estimation \cite{Gonzalez2021} method starts by computing a quantum feature map $x_i \rightarrow \ket{\psi_i}$ based on random Fourier features (RFF) \cite{Rahimi2009RandomMachines}, over a training data set $X = \{x_i\}_{i=1, \cdots, N}$, where $\ket{\psi_i}$ is a normalized vector. Then a training density matrix $\rho$ is constructed as a maximally mixed state of all the $N$ training samples,

\begin{equation}
    \rho = \frac{1}{N}\sum_{i=1}^{N}\ket{\psi_{i}}\bra{\psi_{i}}. \label{eq:3}
\end{equation}

To predict the density of a testing sample $x \rightarrow \ket{\psi}$, the expected value of the sample with the training density matrix is computed, 

\begin{equation}
	\bra{\psi}\rho\ket{\psi}. \label{eq:4}
\end{equation}

The DMKDE in conjunction with RFF works as a non-parametric density estimator, which can approximate probability density functions. 

\subsection{Density Matrix Kernel Density Classification (DMKDC)} \label{dmkdc_section}

The DMKDE can be used for classification as in the Density Matrix Kernel Density Classification method (DMKDC) \cite{Gonzalez2021}. This algorithm creates a quantum feature map of the training and testing samples $x_i \rightarrow \ket{\psi_i}$, some possible quantum feature maps are based on RFF, and soft-max encoding, as presented in \cite{doi:10.7566/JPSJ.90.044002}. Then for each class $j \in \{0, \cdots, D-1\}$, it computes a training density matrix $\rho_j$. The relative frequency, also called prior, of the training samples per class is computed, $\pi_j = N_j/N$, with $N$ the total number of training data points, and $N_j$ the number of training samples of class $j$. The probability $P_j$ of a testing sample $x \rightarrow \ket{\psi}$ to belong to class $j$, would be given by, 

\begin{equation}
P_j = \frac{\pi_j\bra{\psi}\rho_j\ket{\psi}}{\sum_{k=0}^{D-1}\pi_k\bra{\psi}\rho_k\ket{\psi}}. \label{eq:5}
\end{equation}

These density matrices $\rho_j$ can be trained as an average mixed state of the training samples of each class (see equation \ref{eq:3}), or by stochastic gradient descent, which looks for the optimal parameters of the spectral decomposition of the density matrices, $\rho_j = U_j\Lambda_j U_j^\dagger$ with the training data points, using a categorical cross-entropy loss function, see more details in \cite{Gonzalez2021}.
  
\subsection{QuantumSkynet and high-dimensional quantum gates}

\begin{figure}[H]
    \centering
    \subfloat[\centering ]{\includegraphics[scale=0.28]{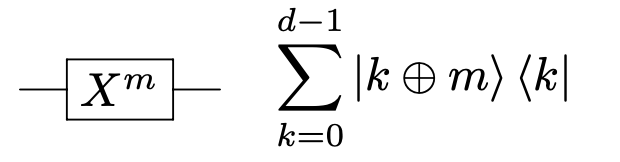} 
    \label{fig:Fig1a}}%
    \subfloat[\centering]{\includegraphics[scale=0.28]{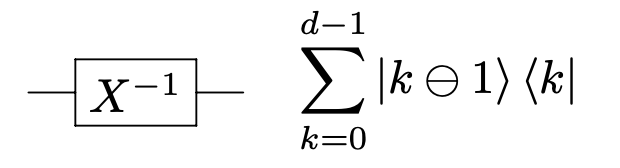} 
    \label{fig:Fig1b}}%
    \caption{\textbf{a} High-dimensional gate $X^{m}$. \textbf{b} High-dimensional gate  $X^{-1}$.}%
    \label{fig:Fig1}%
\end{figure}

QuantumSkynet \cite{giraldocarvajal2021quantumskynet} is a high dimensional quantum computing simulator, that allows to implement high-dimensional quantum algorithms in a cloud-based environment. To simulate the quantum circuits related to this project, the following gates were simulated using QuantumSkynet:

1. The single-qudit gate $X^m$ (see Fig.~\ref{fig:Fig1a}), which is a generalized version of the qubit-based $X$ gate for $d$ dimensions and raised to any $m$ exponent. A particular case of this gate $X$ is when the exponent $m$ is equal to $-1$ (see Fig.~\ref{fig:Fig1b}).



 In Fig.~\ref{fig:Fig1}, $\oplus$ and $\ominus$ stand for summation and subtraction modulo $d$, respectively. 
 
 The result of applying the $X^{m}$ to the canonical basis is $X^m\ket{i}=\ket{i+m}$. In particular, $X^{-1}\ket{i}=\ket{i-1}$.

\begin{figure}[H]
    \centering
    \includegraphics[scale=0.28]{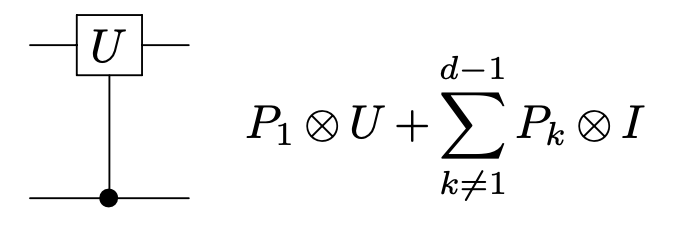}
    \caption{High-dimensional control gate $CU$.}
    \label{fig:Fig3}
\end{figure}

2. The control gate $CU$ (See Fig.~\ref{fig:Fig3}). This control gate applies an arbitrary unitary matrix U only when the control qudit takes the value $\ket{1}$.

Here, $P_k$ is considered as the projection operator equals to $\ket{k}\bra{k}$. For an arbitrary state which results from the tensor product of two qudits, the CU gate does the following transformation,

\begin{equation}
    CU\big(\sum_{i=0}^{d-1}a_{i}\ket{i}\otimes\sum_{j=0}^{d-1}b_{j}\ket{j}\big)=\sum_{\{i: i\ne1\}}a_{i}\ket{i}\otimes\sum_{j=0}^{d-1}b_{j}\ket{j}+ a_1\ket{1}\otimes U(\sum_{j=0}^{d-1}b_{j}\ket{j}), \notag
\end{equation}

with the first qudit as the control and the second qudit as the target.

\begin{figure}[H]
    \centering
    \includegraphics[scale=0.28]{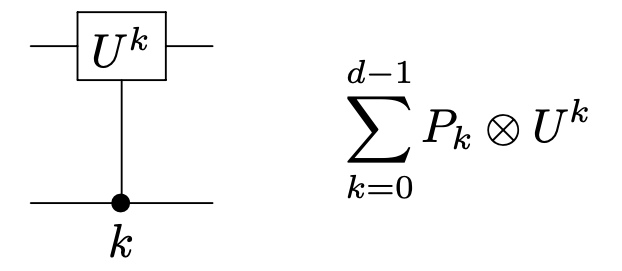}
    \caption{High-dimensional generalized control gate $CU^k$.}
    \label{fig:Fig4}
\end{figure}

3. The generalized controlled gate $CU^k$ (See Fig.~\ref{fig:Fig4}). This gate applies the gate $U^{k}$ to the target qudit, when the control qudit is in state $\ket{k}$, for each possible state of the canonical basis.

One case of interest is when U is equal to $X^{-1}$. In this case, the gate $C(X^{-1})^k$ does the following transformation to an arbitrary two qudit state,

\begin{equation}
    C(X^{-1})^k\big(\sum_{(i,j)}a_{ij}\ket{ij}\big)=\sum_{(i,j)}a_{ij}\ket{i(j-i)}, 
    \label{eq:6}
\end{equation}

where the control is the first qudit and the target is the second qudit.

The $CU^k$ gate can be obtained by a series of $CU$ and $X^{m}$ gates, by the same way as the multiplexer gate \cite{khan2006synthesis}.

\section{Quantum Measurement Classification with Qudits}

The implementation of the algorithms DMKDE and DMKDC requires three phases: (i) Quantum State Preparation, (ii) Training Phase, and (iii) Prediction Phase. The first two steps were computed in a classical computer, with the Tensorflow implementations of the algorithms \cite{Gonzalez2021}, while the prediction phase was simulated in the high dimensional quantum computer simulator QuantumSkynet \cite{giraldocarvajal2021quantumskynet}.

The steps of the DMKDE and DMKDC implementations are:

\begin{enumerate}
	\item Quantum State Preparation: Apply a suitable quantum feature map to the train and test data sets.
	\item Training Phase: Construct the matrices $\rho_j$, one for each class, as a mixed state of the training quantum states, see equation \ref{eq:3}, compute the priors $\pi_j$ of each class, and, calculate the spectral decomposition of these density matrices $\rho_{j} = U_j \Lambda_j U_j^\dag$ (in the DMKDE method there is only one class). 
	\item Prediction Phase: Apply the proposed quantum circuit to make the prediction on each quantum state of the test data set, see equations \ref{eq:4}, \ref{eq:5}.
\end{enumerate}

\begin{figure}[H]
    \centering
    \includegraphics[width=\textwidth]{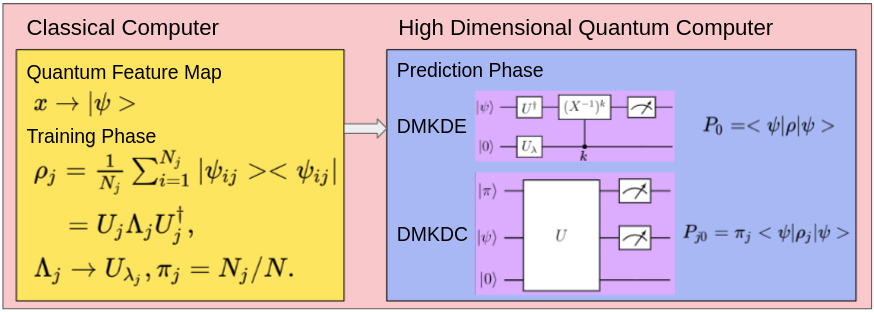}
    \caption{Qudit-based implementation of DMKDE and DMKDC methods. The quantum feature map and the training were performed in a classical computer, while the prediction was done in a high-dimensional quantum computer simulator.}
    \label{general_diagram}
\end{figure}

The main contribution of this article is to show a concrete implementation of the prediction phase of DMKDE and DMKDC as quantum circuits that can be run in a high-dimensional quantum computer. Next, we will present the details of these quantum protocols.

\subsection{Initial comments of the DMKDE and DMKDC quantum circuits}

To implement the prediction phase of DMKDE in a quantum computer with qudits, we should notice that it is equivalent to finding the expected value of a quantum state $\ket{\psi} \in \mathbb{C}^d$ with a training hermitian matrix $\rho$, see equation \ref{eq:4}, therefore, we can apply a spectral decomposition of $\rho$,

\begin{equation}
\bra{\psi}\rho\ket{\psi} = \bra{\psi}U\Lambda U^{\dag}\ket{\psi},
\end{equation}

with $\Lambda$ a diagonal matrix with $\text{Tr}(\Lambda)=1$, and $U$ a unitary matrix. But, $\Lambda = \sum_{i=0}^{d-1}\lambda_i\ket{i}\bra{i}$, then, 
\begin{equation}
 \bra{\psi}\rho\ket{\psi} =   \bra{\psi}U\Big(\sum_{i=0}^{d-1}\lambda_i\ket{i}\bra{i}\Big) U^{\dag}\ket{\psi} = \sum_{i=0}^{d-1}\lambda_i\norm{\bra{i}U^\dag\ket{\psi}}^2. \label{eq:7}
\end{equation}

This form of the DMKDE can be implemented in a high-dimensional quantum computer.

In addition, it is worth mentioning that the DMKDE quantum circuit starts by a assuming we have a suitable quantum feature map of the testing sample $x \rightarrow \ket{\psi}$. The quantum feature map might be based  on random Fourier features or soft-max encoding, as presented in \cite{doi:10.7566/JPSJ.90.044002}. Also, we should have a training density matrix $\rho$ and the resulting matrices of its spectral decomposition $U$ and $\Lambda$. In addition, based on the matrix of eigenvalues $\Lambda$, we require the unitary transformation $U_\lambda$, which satisfies that $U_\lambda \ket{0} = \ket{\lambda}$, where, 

\begin{equation}
	\ket{\lambda} = \sum_{i=0}^{d-1}\sqrt{\lambda_i}\ket{i}. \label{eq:8}
\end{equation}

That is the state $\ket{\lambda}$ is a quantum state which encodes the eigenvalues of the spectral decomposition of $\rho$.

The same previous arguments can be extended to the DMKDC method, but instead of having only one training density matrix $\rho$, in the DMKDC we should have a density matrix $\rho_j$ for each class of the data set. 

These previous steps, i.e., the quantum feature map, the calculation of the training density matrices $\rho_j$, and their spectral decompositions, were done in a classical computer, following the tensorflow implementation of the method \cite{Gonzalez2021}. 

\subsection{The DMKDE quantum circuit}

\begin{figure}[H]
    \centering
    \includegraphics[scale=0.150]{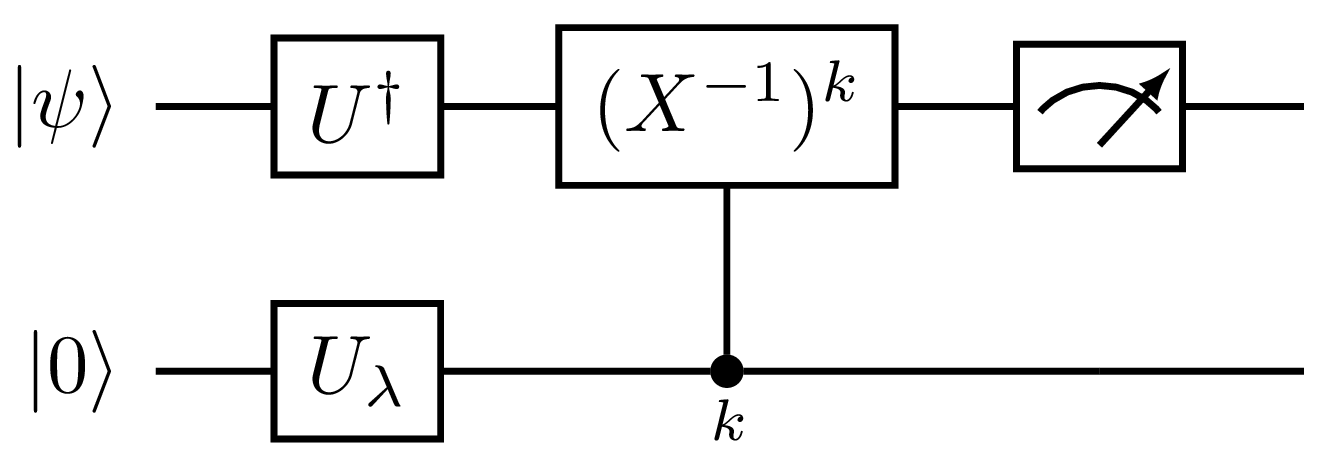}
    \caption{DMDKE high-dimensional quantum circuit.}
    \label{qmkde_circuit}
\end{figure}

The diagram of the DMKDE quantum circuit is presented in figure \ref{qmkde_circuit}. It requires two qudits each of dimension $d$. From the classical computer, we obtain the quantum feature map of the input sample $\ket{\psi} \in \mathbb{C}^d$, the matrix $U$ of eigenvectors of the spectral decomposition of $\rho$, and the unitary matrix $U_\lambda$, which satisfies that, $U_\lambda \ket{0} = \ket{\lambda}$, see equation \ref{eq:8}.

The circuit is initialized with $\ket{\psi}$ in the first qudit, and $\ket{0}$, in the second qudit, 
\begin{equation}
	\ket{\psi} \otimes \ket{0},
	\label{eq:9}
\end{equation}

being the first qudit the leftmost in the equation \ref{eq:9}.

Then, we apply the unitary transformation $U^\dagger$ to the first qudit and $U_\lambda $ to the second qudit, 

\begin{equation}
	U^\dagger\ket{\psi} \otimes U_\lambda \ket{0}= U^\dagger\ket{\psi} \otimes \sum_{j=0}^{d-1}\sqrt{\lambda_j}\ket{j}.
\end{equation}

We can write $U^\dagger\ket{\psi} = \sum_{i=0}^{d-1} a_i\ket{i}$, where $\norm{a_i}^2=\norm{\bra{i}U^\dag\ket{\psi}}^2$. Hence, the coefficients $\norm{a_i}^2$ are the probabilities to measure $U^\dagger\ket{\psi}$ in the canonical basis. Then we can write, 

\begin{align}
U^\dagger\ket{\psi} \otimes U_\lambda \ket{0} &=  \sum_{i=0}^{d-1} a_i\ket{i} \otimes \sum_{j=0}^{d-1}\sqrt{\lambda_j}\ket{j} \notag \\
&=  \sum_{i=0}^{d-1} a_i \sqrt{\lambda_i}\ket{ii} + \sum_{\{(i, j): i\ne j\}} a_i \sqrt{\lambda_j}\ket{ij}. \notag
\end{align}

We can then apply the generalized control gate $C(X^{-1})^k$ with control the second qudit and target the first qudit. Which based on equation \ref{eq:6} results in, 

\begin{equation}
	C(X^{-1})^k(U^\dagger\ket{\psi} \otimes U_\lambda \ket{0}) = \sum_{i=0}^{d-1} a_i \sqrt{\lambda_i}\ket{0i} + \sum_{\{(i, j): i\ne j\}} a_i \sqrt{\lambda_j}\ket{(i-j)j}.
\end{equation}

Finally, by measuring the first qudit the probability of the state $\ket{0}$ is, 
\begin{equation}
P_0 = \sum_{i=0}^{d-1} \norm{a_i}^2 \lambda_i =
\sum_{i=0}^{d-1}\lambda_i\norm{\bra{i}U^\dag\ket{\psi}}^2 = \bra{\psi}\rho\ket{\psi},
\end{equation}

see equation \ref{eq:7}.

\subsection{The DMKDC quantum circuit}

\begin{figure}[H]
    \centering
    \includegraphics[width=\textwidth]{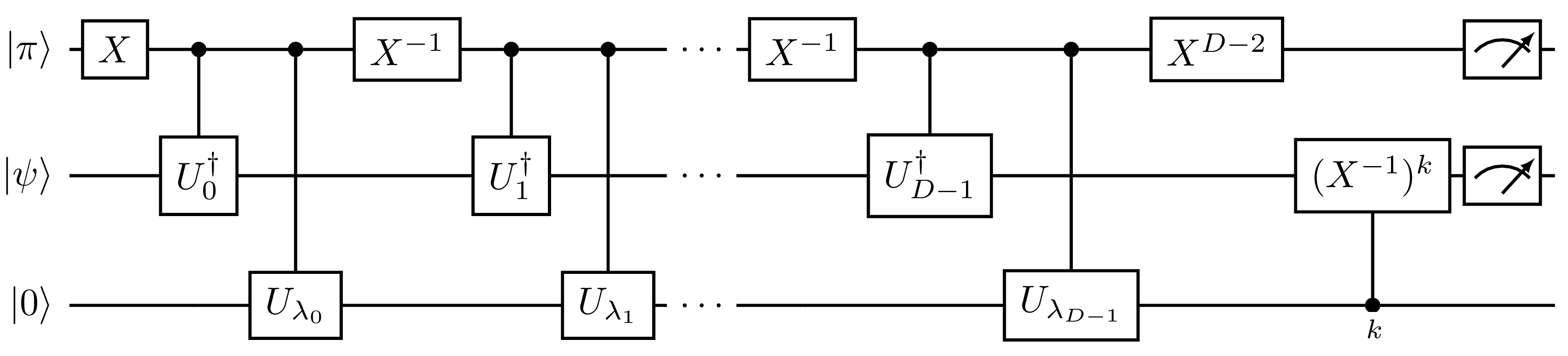}
    \caption{DMKDC high-dimensional quantum circuit.}
    \label{dmkdc_circuit}
\end{figure}

The proposed high-dimensional quantum circuit of the DMKDC is presented in figure \ref{dmkdc_circuit}. As mentioned in section \ref{dmkdc_section}, the DMKDC algorithm requires $D$ density matrices $\rho_j$, one for each class, whose spectral decompositions are given by $\rho_j= U_j\Lambda_jU_j^\dag$. These training density matrices are computed by equation \ref{eq:3} in a classical computer.

The method requires three qudits in $\mathbb{C}^d$, assuming $d \geq D$. The first qudit encodes each of the $D$ classification classes, and the relative frequencies of the training data per class $\pi_j$ (priors). The second qudit serves for two purposes, the input sample to be classified $x\rightarrow \ket{\psi}$ in $\mathbb{C}^d$, at which a suitable quantum feature map has been applied in advance, and for the matrices of eigenvectors $U_j$, each class has a unitary matrix of eigenvectors. In addition, the third qudit is responsible for the eigenvalues of each of the trained density matrices, this can be achieved by the rotation matrices $U_{\lambda j}$, which have the property that $U_{\lambda_j}\ket{0} = \ket{\lambda_j} = \sum_{i=0}^{d-1}\sqrt{\lambda_{ij}}\ket{i}$, where $\lambda_{ij}$ is the $i^{\text{th}}$ eigenvalue of the density matrix $\rho_j$.

The circuit is initialized by $\ket{\pi} = \sum_{j=0}^{D-1}\sqrt{\pi_j}\ket{j}$ in the first qudit, by $\ket{\psi}$ in the second qudit, and by $\ket{0}$ in the third qudit, 

\begin{equation}
	\ket{\pi} \otimes \ket{\psi} \otimes \ket{0} = \Big(\sum_{j=0}^{D-1}\sqrt{\pi_j}\ket{j}\Big) \otimes \ket{\psi} \otimes \ket{0},
\end{equation}

Since the first qudit encodes the classes of the algorithm, the circuit works by changing the control class of the D classes with $X^n$ gates and applying control $CU$ gates for the eigenvalues and eigenvectors of each of the classes. Hence, we first apply a $X$ gate to the first qudit, to change the control class from class 1 to class 0, the result is, 

\begin{equation}
	X(\ket{\pi}) \otimes \ket{\psi} \otimes \ket{0} = \Big(\sum_{j=0}^{D-1}\sqrt{\pi_j}\ket{j + 1}\Big) \otimes \ket{\psi} \otimes \ket{0}.
\end{equation}

We then apply the $CU^\dag_0$ gate with control qudit the first qudit and target the second qudit, and the $CU_{\lambda_0}$ with control qudit the first qubit and target the third qudit, see figure \ref{dmkdc_circuit}, 

\begin{equation}
	\sqrt{\pi_0}\ket{1}\otimes U_0^\dag\ket{\psi} \otimes \ket{\lambda_0} + \Big(\sum_{j=1}^{D-1}\sqrt{\pi_j}\ket{j + 1}\Big) \otimes \ket{\psi} \otimes \ket{0}.
\end{equation}

To replicate the process for class 1, we change the control class from class 0 to class 1, by applying the gate $X^{-1}$ to the first qudit, the result is, 

\begin{equation}
	\sqrt{\pi_0}\ket{0}\otimes U_0^\dag\ket{\psi} \otimes \ket{\lambda_0} + \Big(\sum_{j=1}^{D-1}\sqrt{\pi_j}\ket{j}\Big) \otimes \ket{\psi} \otimes \ket{0}.
\end{equation}

We apply the $CU^\dag_1$ gate with control qudit the first qudit and target the second qudit, and the $CU_{\lambda_1}$ with control qudit the first qudit and target the third qudit,

\begin{equation}
	\sqrt{\pi_0}\ket{0}\otimes U_0^\dag\ket{\psi} \otimes \ket{\lambda_0} + \sqrt{\pi_1}\ket{1}\otimes U_1^\dag\ket{\psi} \otimes \ket{\lambda_1} + \Big(\sum_{j=2}^{D-1}\sqrt{\pi_j}\ket{j}\Big) \otimes \ket{\psi} \otimes \ket{0}.
\end{equation}

We then extend the same block of the class 1 to the restating $D-2$ classes, leaving the class $D-1$ as the control class at the end. Hence, we would have that, 

\begin{equation}
	\sum_{j=0}^{D-1}\big(\sqrt{\pi_j}\ket{j - (D-2)} \otimes U_j^\dag\ket{\psi} \otimes \ket{\lambda_j}\big).
\end{equation}

To restore the class j to the corresponding $\ket{j}$, we apply the gate $X^{D-2}$ to the first qudit,

\begin{equation}
	\sum_{j=0}^{D-1}\big(\sqrt{\pi_j}\ket{j} \otimes U_j^\dag\ket{\psi} \otimes \ket{\lambda_j}\big).
\end{equation}

As with the DMKDE, we can write, 

\begin{align}
U_j^\dagger\ket{\psi} \otimes \ket{\lambda_j} &= \sum_{l=0}^{d-1} a_{lj}\ket{l} \otimes \sum_{m=0}^{d-1}\sqrt{\lambda_{mj}}\ket{m} \notag \\
&= \sum_{l=0}^{d-1} a_{lj} \sqrt{\lambda_{lj}}\ket{ll} + \sum_{\{(l, m): l\ne m\}} a_{lj} \sqrt{\lambda_{mj}}\ket{lm},
\end{align}

 where $\norm{a_{lj}}^2=\norm{\bra{l}U_j^\dag\ket{\psi}}^2$.

Finally, by the same argument of the DMKDE, we would have that by applying the generalized $C(X^{-1})^k$ with control qudit the third qudit and target the second qudit, the circuit leads,

\begin{equation}
	\sum_{j=0}^{D-1}\Big(\sqrt{\pi_j}\ket{j} \otimes \big( \sum_{l=0}^{d-1} a_{lj} \sqrt{\lambda_{lj}}\ket{0l} + \sum_{\{(l, m): l\ne m\}} a_{lj} \sqrt{\lambda_{mj}}\ket{(l-m)m}\big)\Big).
\end{equation}

The desired result is achieved by measuring the amplitudes of the first two qudits. We would have that, 

\begin{equation}
P_{j0} =
\pi_j\sum_{l=0}^{d-1}\lambda_{lj}\norm{\bra{l}U_j^\dag\ket{\psi}}^2 =\pi_j \bra{\psi}\rho_j\ket{\psi},
\end{equation}

see equation \ref{eq:7}. The sample $\ket{\psi}$ will be classified based on,
\begin{equation}
 \max\limits_{j}(\pi_j \bra{\psi}\rho_j\ket{\psi}).
\end{equation}

\section{Results}

We applied the DMKDE and DMKDC circuits to two data sets. We found that the results of the quantum circuits simulated in the high-dimensional quantum simulator QuantumSkynet mimic the results obtained in the Tensorflow implementation of the DMKDE and DMKDC by Gonzalez et. al. \cite{Gonzalez2021}.

For the DMKDE method, we used a 1-D synthetic data set, The training data set corresponded to 1000 points sampled from the linear combination of two Gaussian functions, and there were 1000 equally spaced data points for testing as in \cite{Gonzalez2021}. In this article, they show that DMKDE in combination with random Fourier features (RFF) can approximate any probability density function (pdf). 

\begin{figure}[H]
    \centering
    \includegraphics[scale=0.5]{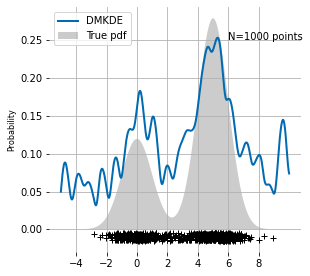}
    \caption{Predictions of the DMKDE quantum circuit with 18-dits of RFF. The DMKDE can approximate probability density functions with high enough RFF.}
    \label{dmkde_model}
\end{figure}

For the implementation of the DMKDE in the high-dimensional quantum computer simulator, we applied a quantum feature map based on RFF  to the raw data with 18 components. Figure \ref{dmkde_model} shows the results of the predictions of the DMKDE quantum circuit. Even though more RFF components would create a better approximation of the pdf, we were restricted by the maximum number of qudit components allowed by the quantum computer simulator.

\begin{figure}
    \centering
    \subfloat[\centering Moons ]{\includegraphics[width=5.8cm]{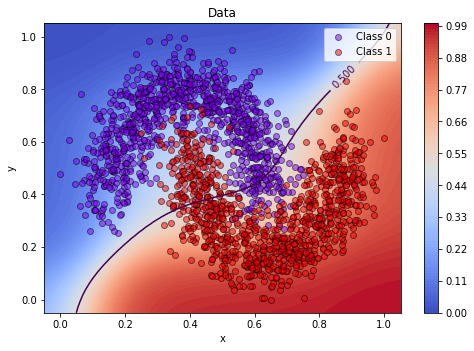} }%
    \subfloat[\centering Circles]{\includegraphics[width=5.8cm]{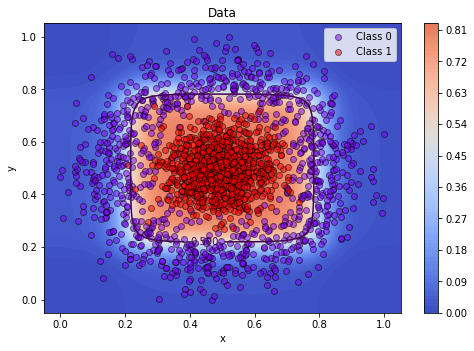} }%
    \caption{Binary predictions of the DMKDC circuit with 9-dits on moons and circles.}%
    \label{circles_moons}%
\end{figure}

Furthermore, we classified two two-dimensional binary data sets of moons and circles to test the DMKDC quantum circuit, see Figure \ref{circles_moons}. There were 1340 samples for training and 660 for testing in each data set. A quantum feature map based on the softmax encoding \cite{doi:10.7566/JPSJ.90.044002} was applied to each data set, resulting in quantum features of 9 dimensions. Therefore, the quantum circuit was constructed with qudits of 9 components. In Figure \ref{circles_moons}, we show the classification boundaries, and the regions with higher probabilities to be classified as either class. We obtained an accuracy of 86.66\% on the test data set of moons, and of 83.63\% on the test data set of circles. The results of the predictions with the high-dimensional quantum circuit are consistent with the Tensorflow implementation of the DMKDC.


Higher accuracies in the DMKDC quantum circuit would have been obtained, if we had used a quantum feature map based on RFF with a higher number of components, and if we had implemented a quantum circuit learning, in which the weights of the trained density matrices are learned by methods like back-propagation.

\section{Conclusions}

In the present article, we showed how to implement in a high-dimensional quantum computer the prediction phase of the quantum-inspired machine learning methods Density Matrix Kernel Density Estimation (DMKDE), and Density Matrix Kernel Density Classification (DMKDC) proposed by Gonzalez et. al. \cite{Gonzalez2021}. The DMKDE and DMKDC quantum circuits were simulated in the qudit-based quantum computer simulator QuantumSkynet \cite{giraldocarvajal2021quantumskynet}. The DMKDE quantum protocol can be extended to compute the expected value of a density matrix with qudits.

Much work is to be done on reducing the complexity of the quantum computer simulator to apply the DMKDC method to more realistic machine learning problems like MNIST, and to improve the capability of the DMKDE to approximate probability density functions by increasing the number of random Fourier features. Furthermore, this framework opens up the possibility to implement the DMKDE and DMKDC algorithms with stochastic-gradient descent in which the weights of the trained density matrices are learned by some optimization procedure, improving the performance of the density estimation and classification.




%
%


\bibliographystyle{unsrtnat}

\bibliography{references}

\begin{thebibliography}{24}
\providecommand{\natexlab}[1]{#1}
\providecommand{\url}[1]{\texttt{#1}}
\expandafter\ifx\csname urlstyle\endcsname\relax
  \providecommand{\doi}[1]{doi: #1}\else
  \providecommand{\doi}{doi: \begingroup \urlstyle{rm}\Url}\fi

\bibitem[Arute et~al.(2019)Arute, Arya, Babbush, Bacon, Bardin, Barends,
  Biswas, Boixo, Brandao, Buell, Burkett, Chen, Chen, Chiaro, Collins,
  Courtney, Dunsworth, Farhi, Foxen, Fowler, Gidney, Giustina, Graff, Guerin,
  Habegger, Harrigan, Hartmann, Ho, Hoffmann, Huang, Humble, Isakov, Jeffrey,
  Jiang, Kafri, Kechedzhi, Kelly, Klimov, Knysh, Korotkov, Kostritsa, Landhuis,
  Lindmark, Lucero, Lyakh, Mandr{\`{a}}, Mcclean, Mcewen, Megrant, Mi,
  Michielsen, Mohseni, Mutus, Naaman, Neeley, Neill, Niu, Ostby, Petukhov,
  Platt, Quintana, Rieffel, Roushan, Rubin, Sank, Satzinger, Smelyanskiy, Sung,
  Trevithick, Vainsencher, Villalonga, White, Yao, Yeh, Zalcman, Neven, and
  Martinis]{Arute2019QuantumProcessor}
Frank Arute, Kunal Arya, Ryan Babbush, Dave Bacon, Joseph~C Bardin, Rami
  Barends, Rupak Biswas, Sergio Boixo, Fernando G S~L Brandao, David~A Buell,
  Brian Burkett, Yu~Chen, Zijun Chen, Ben Chiaro, Roberto Collins, William
  Courtney, Andrew Dunsworth, Edward Farhi, Brooks Foxen, Austin Fowler, Craig
  Gidney, Marissa Giustina, Rob Graff, Keith Guerin, Steve Habegger, Matthew~P
  Harrigan, Michael~J Hartmann, Alan Ho, Markus Hoffmann, Trent Huang, Travis~S
  Humble, Sergei~V Isakov, Evan Jeffrey, Zhang Jiang, Dvir Kafri, Kostyantyn
  Kechedzhi, Julian Kelly, Paul~V Klimov, Sergey Knysh, Alexander Korotkov,
  Fedor Kostritsa, David Landhuis, Mike Lindmark, Erik Lucero, Dmitry Lyakh,
  Salvatore Mandr{\`{a}}, Jarrod~R Mcclean, Matthew Mcewen, Anthony Megrant,
  Xiao Mi, Kristel Michielsen, Masoud Mohseni, Josh Mutus, Ofer Naaman, Matthew
  Neeley, Charles Neill, Murphy~Yuezhen Niu, Eric Ostby, Andre Petukhov, John~C
  Platt, Chris Quintana, Eleanor~G Rieffel, Pedram Roushan, Nicholas~C Rubin,
  Daniel Sank, Kevin~J Satzinger, Vadim Smelyanskiy, Kevin~J Sung, Matthew~D
  Trevithick, Amit Vainsencher, Benjamin Villalonga, Theodore White, Z~Jamie
  Yao, Ping Yeh, Adam Zalcman, Hartmut Neven, and John~M Martinis.
\newblock {Quantum supremacy using a programmable superconducting processor}.
\newblock \emph{Nature}, 574:\penalty0 505, 2019.
\newblock \doi{10.1038/s41586-019-1666-5}.
\newblock URL \url{https://doi.org/10.1038/s41586-019-1666-5}.

\bibitem[Schaeff et~al.(2015)Schaeff, Polster, Huber, Ramelow, and
  Zeilinger]{Schaeff2015ExperimentalOptics}
Christoph Schaeff, Robert Polster, Marcus Huber, Sven Ramelow, and Anton
  Zeilinger.
\newblock {Experimental access to higher-dimensional entangled quantum systems
  using integrated optics}.
\newblock \emph{Optica}, 2\penalty0 (6):\penalty0 523, 6 2015.
\newblock ISSN 2334-2536.
\newblock \doi{10.1364/optica.2.000523}.
\newblock URL \url{http://dx.doi.org/10.1364/OPTICA.2.000523}.

\bibitem[Carolan et~al.(2015)Carolan, Harrold, Sparrow,
  Mart{\'{i}}n-L{\'{o}}pez, Russell, Silverstone, Shadbolt, Matsuda, Oguma,
  Itoh, Marshall, Thompson, Matthews, Hashimoto, O'Brien, and
  Laing]{Carolan2015UniversalOptics}
Jacques Carolan, Christopher Harrold, Chris Sparrow, Enrique
  Mart{\'{i}}n-L{\'{o}}pez, Nicholas~J. Russell, Joshua~W. Silverstone,
  Peter~J. Shadbolt, Nobuyuki Matsuda, Manabu Oguma, Mikitaka Itoh, Graham~D.
  Marshall, Mark~G. Thompson, Jonathan~C.F. Matthews, Toshikazu Hashimoto,
  Jeremy~L. O'Brien, and Anthony Laing.
\newblock {Universal linear optics}.
\newblock \emph{Science}, 349\penalty0 (6249):\penalty0 711--716, 8 2015.
\newblock ISSN 10959203.
\newblock \doi{10.1126/science.aab3642}.
\newblock URL \url{https://pubmed.ncbi.nlm.nih.gov/26160375/}.

\bibitem[Sit et~al.(2017)Sit, Bouchard, Fickler, Gagnon-Bischoff, Larocque,
  Heshami, Elser, Peuntinger, G{\"{u}}nthner, Heim, Marquardt, Leuchs, Boyd,
  and Karimi]{Sit2017High-dimensionalPhotons}
Alicia Sit, Frédéric Bouchard, Robert Fickler, Jérémie Gagnon-Bischoff,
  Hugo Larocque, Khabat Heshami, Dominique Elser, Christian Peuntinger, Kevin
  G{\"{u}}nthner, Bettina Heim, Christoph Marquardt, Gerd Leuchs, Robert~W.
  Boyd, and Ebrahim Karimi.
\newblock {High-dimensional intracity quantum cryptography with structured
  photons}.
\newblock \emph{Optica}, 4\penalty0 (9):\penalty0 1006, 9 2017.
\newblock ISSN 2334-2536.
\newblock \doi{10.1364/optica.4.001006}.
\newblock URL \url{https://doi.org/10.1364/OPTICA.4.001006}.

\bibitem[Klimov et~al.(2003)Klimov, Guzm{\'{a}}n, Retamal, and
  Saavedra]{Klimov2003QutritIons}
A.~B. Klimov, R.~Guzm{\'{a}}n, J.~C. Retamal, and C.~Saavedra.
\newblock {Qutrit quantum computer with trapped ions}.
\newblock \emph{Physical Review A - Atomic, Molecular, and Optical Physics},
  67\penalty0 (6):\penalty0 7, 6 2003.
\newblock ISSN 10941622.
\newblock \doi{10.1103/PhysRevA.67.062313}.
\newblock URL
  \url{https://journals.aps.org/pra/abstract/10.1103/PhysRevA.67.062313}.

\bibitem[Gedik et~al.(2015)Gedik, Silva, {\c{C}}akmak, Karpat, Vidoto,
  Soares-Pinto, DeAzevedo, and Fanchini]{Gedik2015ComputationalQudit}
Z.~Gedik, I.~A. Silva, B.~{\c{C}}akmak, G.~Karpat, E.~L.G. Vidoto, D.~O.
  Soares-Pinto, E.~R. DeAzevedo, and F.~F. Fanchini.
\newblock {Computational speed-up with a single qudit}.
\newblock \emph{Scientific Reports}, 5\penalty0 (1):\penalty0 14671, 10 2015.
\newblock ISSN 20452322.
\newblock \doi{10.1038/srep14671}.
\newblock URL \url{www.nature.com/scientificreports/}.

\bibitem[Moreno-Pineda et~al.(2018)Moreno-Pineda, Godfrin, Balestro,
  Wernsdorfer, and Ruben]{Moreno-Pineda2018MolecularAlgorithms}
Eufemio Moreno-Pineda, Clément Godfrin, Franck Balestro, Wolfgang Wernsdorfer,
  and Mario Ruben.
\newblock {Molecular spin qudits for quantum algorithms}, 1 2018.
\newblock ISSN 14604744.
\newblock URL
  \url{https://pubs.rsc.org/en/content/articlehtml/2018/cs/c5cs00933b
  https://pubs.rsc.org/en/content/articlelanding/2018/cs/c5cs00933b}.

\bibitem[Cozzolino et~al.(2019)Cozzolino, Da~Lio, Bacco, and
  Oxenl{\o}we]{Cozzolino2019High-dimensionalChallenges}
Daniele Cozzolino, Beatrice Da~Lio, Davide Bacco, and Leif~Katsuo Oxenl{\o}we.
\newblock {High-dimensional quantum communication: benefits, progress and
  future challenges}.
\newblock Technical report, 2019.

\bibitem[Sheridan and Scarani(2010)]{Sheridan2010SecuritySystems}
Lana Sheridan and Valerio Scarani.
\newblock {Security proof for quantum key distribution using qudit systems}.
\newblock \emph{Physical Review A - Atomic, Molecular, and Optical Physics},
  82\penalty0 (3):\penalty0 030301, 9 2010.
\newblock ISSN 10502947.
\newblock \doi{10.1103/PhysRevA.82.030301}.
\newblock URL
  \url{https://journals.aps.org/pra/abstract/10.1103/PhysRevA.82.030301}.

\bibitem[Rebentrost et~al.(2014)Rebentrost, Mohseni, and
  Lloyd]{Rebentrost2014QuantumClassification}
Patrick Rebentrost, Masoud Mohseni, and Seth Lloyd.
\newblock {Quantum support vector machine for big data classification}.
\newblock \emph{Physical Review Letters}, 113\penalty0 (3), 9 2014.
\newblock ISSN 10797114.
\newblock \doi{10.1103/PhysRevLett.113.130503}.

\bibitem[Wiebe et~al.(2014)Wiebe, Kapoor, and Svore]{Wiebe2014QuantumLearning}
Nathan Wiebe, Ashish Kapoor, and Krysta~M. Svore.
\newblock {Quantum Deep Learning}.
\newblock 12 2014.

\bibitem[Lu and Braunstein(2014)]{Lu2014QuantumClassifier}
Songfeng Lu and Samuel~L Braunstein.
\newblock {Quantum decision tree classifier}.
\newblock 13:\penalty0 757--770, 2014.
\newblock \doi{10.1007/s11128-013-0687-5}.

\bibitem[Ezhov and Ventura(2000)]{Ezhov2000QuantumNetworks}
Alexandr~A. Ezhov and Dan Ventura.
\newblock {Quantum Neural Networks}.
\newblock pages 213--235. Physica, Heidelberg, 2000.
\newblock \doi{10.1007/978-3-7908-1856-7{\_}11}.
\newblock URL
  \url{https://link.springer.com/chapter/10.1007/978-3-7908-1856-7_11}.

\bibitem[Cong et~al.(2019)Cong, Choi, and Lukin]{Cong2019QuantumNetworks}
Iris Cong, Soonwon Choi, and Mikhail~D. Lukin.
\newblock {Quantum convolutional neural networks}.
\newblock \emph{Nature Physics}, 15\penalty0 (12):\penalty0 1273--1278, 12
  2019.
\newblock ISSN 17452481.
\newblock \doi{10.1038/s41567-019-0648-8}.
\newblock URL \url{https://www.nature.com/articles/s41567-019-0648-8}.

\bibitem[Dallaire-Demers and
  Killoran(2018)]{Dallaire-Demers2018QuantumNetworks}
Pierre-Luc Dallaire-Demers and Nathan Killoran.
\newblock {Quantum generative adversarial networks}.
\newblock Technical report, 2018.

\bibitem[C{\'{a}}rdenas-L{\'{o}}pez et~al.(2018)C{\'{a}}rdenas-L{\'{o}}pez,
  Lamata, Retamal, and Solano]{Cardenas-Lopez2018MultiqubitTechnologies}
F.~A. C{\'{a}}rdenas-L{\'{o}}pez, L.~Lamata, J.~C. Retamal, and E.~Solano.
\newblock {Multiqubit and multilevel quantum reinforcement learning with
  quantum technologies}.
\newblock \emph{PLoS ONE}, 13\penalty0 (7):\penalty0 e0200455, 7 2018.
\newblock ISSN 19326203.
\newblock \doi{10.1371/journal.pone.0200455}.
\newblock URL \url{https://doi.org/10.1371/journal.pone.0200455}.

\bibitem[Diep(2020)]{Diep2020SomeNetworks}
Do~Ngoc Diep.
\newblock {Some Quantum Neural Networks}.
\newblock \emph{International Journal of Theoretical Physics}, 59\penalty0
  (4):\penalty0 1179--1187, 4 2020.
\newblock ISSN 15729575.
\newblock \doi{10.1007/s10773-020-04397-1}.
\newblock URL
  \url{https://link.springer.com/article/10.1007/s10773-020-04397-1}.

\bibitem[Ricks and Ventura(2004)]{NIPS2003_50525975}
Bob Ricks and Dan Ventura.
\newblock Training a quantum neural network.
\newblock In S.~Thrun, L.~Saul, and B.~Sch\"{o}lkopf, editors, \emph{Advances
  in Neural Information Processing Systems}, volume~16. MIT Press, 2004.
\newblock URL
  \url{https://proceedings.neurips.cc/paper/2003/file/505259756244493872b7709a8a01b536-Paper.pdf}.

\bibitem[Beer et~al.(2019)Beer, Bondarenko, Farrelly, Osborne, Salzmann, and
  Wolf]{Beer2019EfficientNetworks}
Kerstin Beer, Dmytro Bondarenko, Terry Farrelly, Tobias~J. Osborne, Robert
  Salzmann, and Ramona Wolf.
\newblock {Efficient learning for deep quantum neural networks}, 2 2019.
\newblock ISSN 23318422.
\newblock URL \url{https://doi.org/10.1038/s41467-020-14454-2}.

\bibitem[Gonz{\'{a}}lez et~al.(2021)Gonz{\'{a}}lez, Gallego,
  Toledo-Cort{\'{e}}s, and Vargas-Calder{\'{o}}n]{Gonzalez2021}
Fabio~A. Gonz{\'{a}}lez, Alejandro Gallego, Santiago Toledo-Cort{\'{e}}s, and
  Vladimir Vargas-Calder{\'{o}}n.
\newblock {Learning with Density Matrices and Random Features}.
\newblock 2021.
\newblock URL \url{http://arxiv.org/abs/2102.04394}.

\bibitem[Giraldo-Carvajal et~al.(2021)Giraldo-Carvajal, Duque-Ramirez, and
  Jaramillo-Villegas]{giraldocarvajal2021quantumskynet}
Andres Giraldo-Carvajal, Daniel~A. Duque-Ramirez, and Jose~A.
  Jaramillo-Villegas.
\newblock Quantumskynet: A high-dimensional quantum computing simulator, 2021.

\bibitem[Rahimi and Recht(2009)]{Rahimi2009RandomMachines}
Ali Rahimi and Benjamin Recht.
\newblock {Random features for large-scale kernel machines}.
\newblock In \emph{Advances in Neural Information Processing Systems 20 -
  Proceedings of the 2007 Conference}, 2009.
\newblock ISBN 160560352X.

\bibitem[González et~al.(2021)González, Vargas-Calderón, and
  Vinck-Posada]{doi:10.7566/JPSJ.90.044002}
Fabio~A. González, Vladimir Vargas-Calderón, and Herbert Vinck-Posada.
\newblock Classification with quantum measurements.
\newblock \emph{Journal of the Physical Society of Japan}, 90\penalty0
  (4):\penalty0 044002, 2021.
\newblock \doi{10.7566/JPSJ.90.044002}.
\newblock URL \url{https://doi.org/10.7566/JPSJ.90.044002}.

\bibitem[Khan and Perkowski(2006)]{khan2006synthesis}
Faisal~Shah Khan and Marek Perkowski.
\newblock Synthesis of multi-qudit hybrid and d-valued quantum logic circuits
  by decomposition.
\newblock \emph{Theoretical Computer Science}, 367\penalty0 (3):\penalty0
  336--346, 2006.

\end{thebibliography}


\end{document}